\documentclass{article}

\usepackage{arxiv}

\usepackage[utf8]{inputenc} 
\usepackage[T1]{fontenc}    
\usepackage{hyperref}       
\usepackage{url}            
\usepackage{booktabs}       
\usepackage{amsfonts}       
\usepackage{nicefrac}       
\usepackage{microtype}      
\usepackage{lipsum}
\usepackage{comment}

\title{A Scheme to Verify Services with Unboundedly many Clients using NuSMV} 

\author{ 
   S Sheerazuddin \\
    SSN College of Engineering  \\
     Rajiv Gandhi Salai \\
     Kalavakkam, Chennai \\
    	\texttt{sheerazuddins@ssn.edu.in} \\
  \And
   Anand S \\
    SSN College of Engineering  \\
     Rajiv Gandhi Salai \\
     Kalavakkam, Chennai \\
    	\texttt{anand.saminathan1.1@ssn.edu.in} \\
\And
   Anish Badhri R S \\
    SSN College of Engineering  \\
     Rajiv Gandhi Salai \\
     Kalavakkam, Chennai \\
    	\texttt{anishbadhri@ssn.edu.in} \\
}

\newtheorem{dfn}{Definition}[section]

\newtheorem{thm}[dfn]{Theorem}

\newcommand{\RightBox}{{\phantom{a}}\hfill $\Box$ \\}

\newcommand{\diamin}{\Diamond\kern-0.5em{\raisebox{.25ex}{\rm -}}\kern0.175em}

\newcommand{\Longstep}[1]{\mbox{$\stackrel{#1}{\Longrightarrow}$}}

\newcommand{\ldot}{{\rm <}\kern-0.37em{\raisebox{.25ex}{\bf .}}\kern0.375em}

\newcommand{\calI}{\mathcal{I}}

\begin{document}
\maketitle

\begin{abstract} 
We study model checking of client - server systems, where the servers offer 
several types of services that may depend, at any time, on how many clients of specific 
types are active at that time. Since there are unboundedly many clients,
the state space of such systems is infinite, rendering specification
and verification hard. This problem can be circumvented by using a specification
language which has monadic first-order (MFO) sentences closed with standard
temporal modalities. The MFO sentences throw up a bound which can, in turn, be
used to bound the state space of the input client - server system, thereby making
the verification problem decidable. This scheme is implemented using the NuSMV tool.
\end{abstract}


\section{Introduction}
An important abstraction in the study of distributed systems is that of
{\bf client - server systems} \cite{B92}. Rather than consider the system as a
flat parallel composition of processes, we study a hierarchy of server
and clients: the latter send requests to the server and wait for response. 
Such a model is common in the analysis of web services (\cite{Nar}).

Of particular interest in this context are systems where the number of
active clients may be {\em unbounded} at any point of time. These are
systems where the number of active processes at any system state is not 
known at design time but decided only at run time. Though at any point 
of time, only finitely many agents may be participating, there is no 
uniform bound on the number of agents. Design and verification of such 
systems is becoming increasingly important in distributed computing, 
especially is the context of web services. Well known examples of
such services include Loan handling and Travel agency Web Services\cite{BPEL}.

For example, a loan agency may handle clients of different types
depending on the quantum of loans they seek. How much is granted
may well depend on the number of active clients of different types.
In the case of a travel agency, making bookings for one client
may be very different from several of these clients together. 

Since services handling unknown clients need to make decisions based 
upon request patterns that are not pre-decided, they need to conform 
to specific {\bf service policies} that are articulated at design time. 
Due to concurrency and unbounded state information, the design and 
implementation of such services becomes complex and hence subject to 
logical flaws. Thus, there is a need for formal methods in specifying 
service policies and verifying that systems implement them correctly.

{\bf Model checking} (\cite{CGP}) is an important technique in this regard, whereby
the system to be checked is modeled as a finite state system, and properties 
to be verified are expressed as constraints on the possible computations of 
the model. This facilitates algorithmic tools to be employed in verifying 
that the model is indeed correct with respect to those properties. When we 
find violations, we re-examine the finite-state abstraction, leading to a 
finer model, perhaps also refine the specifications and repeat the process. 
Finding good finite state abstractions forms an important part of the approach.

Modeling systems with unboundedly many clients is fraught with difficulties. 
Since we have no bound on the number of active processes, the state space is 
infinite. A finite state abstraction may not suffice since service policies may
involve counting the number of active clients. 



We propose a model for such systems with {\bf passive} clients 
where clients simply send a request and wait for an answer and do not 
interact with the server between the send and receive. The passive client 
systems are modeled as state transition systems which identify clients 
only by the types of service (fixed a priori) that they are associated with. 
Thus, transitions over an alphabet of send-requests and receive-answers 
of different types suffice. The situation is more complex for systems 
with {\bf active} clients, called {\bf session-oriented} services in
the literature \cite{AZTEC}.

Propositional temporal logics, for e.g., $CTL$ \cite{BPM83}, have been extensively 
used  for specifying safety and liveness requirements of reactive systems. Backed 
by a set of tools with theorem proving \cite{Z10} and model checking \cite{McM93} 
capabilities, temporal logic is a natural candidate for specifying service policies. 
In the context of Web Services, they have been extended with mechanisms for 
specifying message exchange between agents. There are several candidate temporal 
logics for message passing systems, for e.g., $m$-LTL \cite{MR}, 
but these work with {\em a priori} fixed number of agents, and for any message, 
the identity of the sender and the receiver are fixed at design time. We need 
to extend such logics with means for referring to agents in some more abstract 
manner (than by name). 

A natural and direct approach to refer to unknown clients is to use logical 
variables: rather than work with atomic propositions $p$, we use monadic 
predicates $p(x)$ to refer to property $p$ being true of client $x$. We can 
then quantify over such $x$ existentially and universally to specify policies 
relating clients. We are thus naturally led to the realm of Monadic First Order 
Temporal Logics ($MFOTL$)\cite{GHR}. It is easily seen that $MFOTL$ is 
expressive enough to frame almost every requirement specification of client - server 
systems of the kind discussed above. Unfortunately $MFOTL$  is undecidable 
\cite{HWZ}, and we need to limit the expressiveness to have a decidable 
verification problem.

Decidable fragments of $MFOTL$ are very few; the {\bf one-variable} 
fragment \cite{HV}, \cite{SZ} and the {\bf monodic} fragment \cite{HWZ} 
are the only nontrivial ones in the literature. The decidability results 
from the fact that there is at most one free variable in the scope of 
temporal modalities. The ``packed'' monodic fragment with equality is 
decidable as well (\cite{H}). \cite{HWZ02} offers similar results for
restricted branching time logics with first-order extensions.
\cite{BL} and \cite{BLb}  extend these techniques for first-order 
temporal epistemic logics. However, the techniques developed are
not automata theoretic, and hence better suited to studying axiom
systems and satisfiability, rather than model checking.

We propose a fragment of monadic temporal logic, named Monadic First-order
Sentential Temporal Logic ($MFSTL$), for which satisfiability and model 
checking are decidable. This language is weak in expressive power but 
reasoning in such a logic is already sufficient to express a broad range 
of service policies in systems with unboundedly many active clients. 

The current paper is a revised version of \cite{She11} where this class of 
systems was introduced and the logic and its satisfiability 
was studied. In this paper, we provide a practical scheme to implement 
specification and verification of such client - server systems using NuSMV.

The work closest to ours is that of LeeTL \cite{MKK17} and Graded-CTL \cite{FNP08}.
Graded-CTL is a strict extension of CTL with graded quantifiers $A$ and $E$ that can be 
used to reason about {\em at least} $k$ or {\em all but} $k$ possible behaviours in a
system. This can, in turn, be used to generate multiple counterexamples in one run
of the model checker. The model checking problem for this logic is solved \cite{FMNPS10} by extending 
the NuSMV tool to accept specifications in graded-CTL. 

LeeTL is an extension of LTL that is used to specify properties of object-based systems
with dynamic object creation. The formulas in LeeTL can be encoded into an extension of 
B\"uchi automata. This enables adaptation of standard LTL model checking algorithm for 
model checking LeeTL formulas.

In our case, the logic $MFSTL$ is expressively equivalent to LTL and the 
model checking problem is solved by implementing a layer over NuSMV which 
accepts specification in $MFSTL$ and encodes it into LTL. The model for system 
is also considerably different from standard Kripke structures, with its 
semantics given by multi-counter automata, and has to be encoded into SMV 
format, by transforming it into a finite-state system using a bound computed 
from the $MFSTL$ specification.


The paper is structured as follows. First, we describe a system model
for server with passive clientele called Service for Passive clientS ($SPS$).
Thereafter, we propose a logic to specify the peoperties of $SPS$-like systems,
namely Monadic First-order Sentential Temporal Logic($MFSTL$). We then study 
model checking of $MFSTL$ against $SPS$ using NuSMV in the penultimate section.
We conclude the paper with a brief discussion on possible future work. 


\section{A Model for Client - Server Systems}
In this section we describe an automaton model for client - server
systems that admits unbounded number of agents. We consider the
simpler case where the clients send requests to the server and
wait for the response (either yes or no). The challenge here is
twofold: to give a finite (and simple) description of an inherently 
infinite-state system--owing to a lack of bound on the number of clients 
and to constrain the model so that it allows decidable reasoning, in 
particular, reachability should be decidable in such a model.
This model was first described in \cite{She11}. 

Fix $CN$, a countable set of {\em client names}. In general, 
this set would be recursively generated using a naming scheme, 
for instance using sequence numbers and time-stamps generated 
by processes. We choose to ignore this structure for the sake 
of technical simplicity. We will use $a,b$ etc.~with or without 
subscripts to denote elements of $CN$.

%
Fix $\Gamma_0$, a finite {\bf service alphabet}. We use $u,v$ 
etc.~to denote elements of $\Gamma_0$, and they are thought of 
as {\bf types of services} provided by a server.  This means 
that when two clients ask for a service of the same type, 
given by an element of $\Gamma_0$ it can tell them apart 
only by their name. We could in fact then insist that server's 
behaviour be identical towards both, but we do not make such an 
assumption, to allow for generality.

We will assume a map $\lambda:CN \to \Gamma_0$ that partitions 
$CN$ into disjoint sets of service types. For $u \in \Gamma_0$,
let $CN_u = \{a \in CN \mid \lambda(a) = u\}$ which is also
assumed to be countable with a pre-defined order. (This is so that we never run out of
client names for any type.) For the rest of the paper we fix 
a triple  $(\Gamma_0, CN, \lambda)$ for the discourse and 
discuss a class of systems and specifications over this alphabet.

An {\em extended alphabet} is a set $\Gamma=\{req_u,ans_u\mid u\in 
\Gamma_0\}\cup\{\tau\}$. These refer to requests for such 
service and answers to such requests, as well as the ``silent'' 
internal action $\tau$.

We define below systems of services that handle passive 
clientele. Servers are modeled as state transition 
systems which identify clients only by the type of service 
they are associated with. Thus, transitions are associated 
with client types rather than client names.
\begin{dfn}
A \textbf{Service for Passive Clients} (SPS) is a tuple 
$A=(S,\delta,I)$ where $S$ is a finite set of states, 
$\delta\subseteq (S\times \Gamma\times S)$ is a server 
transition relation and $I\subseteq S$ is the set of initial 
states.
\end{dfn}
%

Note that an $SPS$ is a finite state description. A transition 
of the form $(s,req_u,s^{\prime})$ refers implicitly to a 
{\bf new} client of type $u$ rather than to any specific 
client name. The meaning of this is provided in the run 
generation mechanism described below.

A {\bf configuration} of an $SPS$ $A$ is a pair $(s,C)$ 
where $s\in S$ and $C$ is a finite subset of $CN$. 
Thus a configuration specifies the control 
state of the server, as well as the finite set of {\em active} 
clients at that configuration. 

Let $\Omega_A$ denote the set of 
all configurations of $A$; note that it is this {\em infinite} 
configuration space that is navigated by {\em runs} of $A$, defined below. A 
configuration $(s,C)$ is said to be {\em initial} if 
$s\in I$ and $C=\emptyset$.

We can extend the transition relation $\delta$ to configurations 
$\Longstep{r} \subseteq (\Omega_A\times \Gamma \times \Omega_A)$ 
as follows: $(s,C)\Longstep{r}(s^{\prime},C^{\prime})$
iff $(s,r,s^{\prime})\in \delta$ and the 
following conditions hold:

\begin{itemize}
\item when $r=\tau$, $C=C^{\prime}$; 
\item when $r=req_u$, $C^{\prime}=C\cup\{a\}$, 
where $a$ is the least element of $CN-C$;
\item when $r=ans_u$, $X=\{a\in C\mid \lambda(a)=u\}\ne \emptyset$, and
$C^{\prime}=C-\{a\}$ where $a$ is the least in the enumeration of $X$. 
\end{itemize}

A \textbf{run} of an $SPS$ $A$ on $r_1 r_2 \ldots r_n \in \Gamma^*$ is a 
sequence of configurations $\rho=c_0c_1\cdots c_n$, where 
$c_0$ is initial, and for all $j>0$, $c_{j-1}\Longstep{r_j} c_j$. 
Let $R_A$ denote the set of all runs of $A$. 

Let $\rho$ be a run as defined above; 
$\rho$ is said to be accepting if $s_n \in F$ where $c_n = (s_n, C_n)$.
(Often we may wish to consider final configurations to be ones
in which $C_n$, the set of active agents, is empty. We do not
impose this constraint here for the sake of generality.)
We define the language accepted by $A$ as: $L(A) = \{w \in 
\Gamma^* \mid$ there exists an accepting run $\rho$ of $A$ on $w\}$.

Note that runs have considerable structure. For instance, 
the configuration space $\Omega_A$ can have an infinite path 
generated by a self-loop of the form $(s,req_x,s)$ in $\delta$ 
which corresponds to an unbounded sequence of service requests 
of a particular type. By studying only finite runs, we miss
some interesting properties but since runs include counter
behaviour, we already have plenty of complexity in the system.


\subsection{Algorithms on $SPS$}
We can view $SPS$ as automaton models for servers with unbounded
clients. Indeed, when we have $k$ client types, such systems
correspond to {\bf $k$-counter automata without zero-tests}.
These automata, long studied as Vector addition systems with
states, and as Petri Nets, have a rich automata theory. Briefly,
every request of type $u$ corresponds to incrementing a counter
of that type, every answer corresponds to a decrement, and
the system gets stuck when it attempts to answer a non-existent
request. Such a correspondence with multi-counter automata is
fortunate: this means that we can simply borrow results from
classical automata theory. Therefore, though the configuration 
space of an $SPS$ is infinite, reachability is decidable, 
according to the celebrated theorem proved independently by
Mayr and Kosaraju. We state the following theorem without proof,
and refer the reader to \cite{HP79}, \cite{KRaju82}, \cite{Mayr81}.
\begin{thm}
The class of languages recognized by $SPS$ are closed under union
and intersection but not under complementation. Given an $SPS$,
checking whether it accepts a non-empty language is decidable.
\end{thm}
%
%

%
%

\section{Monadic First-order Sentential Temporal Logic}
We now describe a logical language to specify and verify SPS-like systems. This
logic was first presented in \cite{She11}. Such a 
language has two mutually exclusive dimensions. One, captured by $MFO$ fragment, 
talking about the plurality of clients asking for a variety of services. The other, 
captured by $LTL$ fragment, talks about the temporal variations of services being 
rendered.  Furthermore, the $MFO$ fragment has to be {\sf multi-sorted} to cover 
the multiplicity of service types. Keeping these issues in mind, we frame a logical 
language, which we call Monadic First-order Sentential Temporal Logic ($MFSTL$), a 
combination of $LTL$ and {\sf multi-sorted} $MFO$. In the case of $LTL$, atomic 
formulae are propositional constants which have no further structure. In $MFSTL$, 
there are two kind of atomic formulae, basic server properties from $P_s$, and 
$MFO$-sentences over client properties $P_c$. Consequently, these formulae are 
interpreted over sequences of $MFO$-structures juxtaposed with $LTL$-models.

At the outset, we fix $\Gamma_0$, a finite set of client types. The set of  {\sf client formulae} are defined over a countable set of atomic {\sf client predicates} $P_c$, which are composed of disjoint predicates $P_c^u$ of type $u$ , for each $u \in \Gamma_0$. Also, let $Var$ be a countable supply of variable symbols and $CN$ be a countable set of client names. $CN$ is divided into disjoint sets of types from $\Gamma_0$ via $\lambda:CN \to \Gamma_0$. Similarly, $Var$ is divided using $\Pi:Var \to \Gamma_0$. We use $x,y$ to denote elements in $Var$ and $a,b$ for elements in $CN$. 

Formally, the set of client formulae $\Phi$ is defined as follows: 

$\alpha,\beta \in\Phi ::= p(x:u), p\in P_c^u\mid x=y,x,y\in Var_u \mid \lnot \alpha \mid \alpha \lor \beta \mid (\exists x:u) \alpha$.

Let $\mathcal{S}_{\Phi}$ be the set of all sentences in $\Phi$, then, the  server formulae are defined as follows:

$\psi \in \Psi ::= lp \in P_s \mid \varphi \in \mathcal{S}_{\Phi} \mid \lnot \psi  \mid \psi_1\lor\psi_2  \mid  X  \psi  \mid \psi U \psi^{\prime}$

Derived modalities $F$ (``eventually'') and $G$ (``always'') are defined in the usual way: $F \alpha \equiv TRUE ~U~ \alpha$ and $G \alpha \equiv \lnot F \lnot \alpha$, where $TRUE$ and $FALSE$ are standard propositional constants.
\subsection{Semantics}
$MFSTL$ is interpreted over sequences of $MFO$ models composed with $LTL$ models.
Formally, a model is a triple $M=(\nu,D,I)$ where
\begin{enumerate}
\item $\nu=\nu_0\nu_1\cdots$, where $\forall i \in \omega$, $\nu_i \subset_{fin} P_s$, gives the local properties of the server at instance $i$,
\item $D=D_0D_1D_2\cdots$, where $\forall i \in \omega$, $D_i= (D_i^u)_{u\in \Gamma_0}$ where $D_i^u\subset_{fin}CN_u$, gives the identity of the clients of each type being served at instance $i$ and 
\item $I=I_0I_1I_2\cdots$, where $\forall i \in \omega$, $I_i=(I_i^u)_{u\in \Gamma_0}$ and $I_i^u:D_i^u\to 2^{P_c^u}$ gives the properties satisfied by each live agent at $i$th instance, in other words, the corresponding states of live agents.
\newline
Alternatively, $I_i^u$ can be given as $I_i^u:D_i^u\times P_c^u \to\{\top,\bot\}$, an equivalent form.
\end{enumerate}
\subsubsection{Satisfiability Relations $\models$, $\models_{\Phi}$}
Let $M=(\nu,D,I)$ be a valid model and $\pi:Var\to CN$ be a partial map consistent with respect to $\lambda$ and $\Pi$. Then, the relations $\models$ and $\models_{\Phi}$ can be defined, via induction over the structure of $\psi$ and $\alpha$, respectively, as follows:
\begin{enumerate}
\item $M,i\models lp$ iff $lp \in \nu_i$.
\item $M,i\models \varphi$ iff $M,\emptyset,i\models_{\Phi} \varphi$.
\item $M,i\models \lnot \psi$ iff $M,i\not\models \psi$.
\item $M,i\models \psi\lor\psi^{\prime}$ iff $M,i\models \psi$ or $M,i\models \psi^{\prime}$.
\item $M,i\models X \psi$ iff $M,i+1\models \psi$.
\item $M,i\models \psi U \psi^{\prime}$ iff $\exists j\ge i$, $M,j\models \psi^{\prime}$ and $\forall i': i\le i^{\prime} <j$, $M,i^{\prime}\models \psi$.
\newline
\item $M,\pi,i \models_{\Phi} p(x:u)$ iff $\pi(x)\in D_i^u$ and $I_i(\pi(x),p)=\top$.
\item $M,\pi,i \models_{\Phi} x=y$ iff $\pi(x)=\pi(y)$.
\item $M,\pi,i \models_{\Phi}\lnot \alpha$ iff $M,\pi,i \not\models_{\Phi}\alpha$.
\item $M,\pi,i \models_{\Phi}\alpha\lor\beta$ iff $M,\pi,i \models_{\Phi}\alpha$ or $M,\pi,i \models_{\Phi}\beta$.
\item $M,\pi,i \models_{\Phi}(\exists x:u)\alpha$ iff $\exists a\in D_i^u$ and $M,\pi[x\mapsto a],i \models_{\Phi}\alpha$. 
\end{enumerate} 
\subsection{Specification Examples Using $MFSTL$}
In this section, we would like to show that our logic $MFSTL$ adequately captures many of the facets of SPS-like systems. We consider the {\sf Loan Approval Web Service} \cite{She11}, and frame specifications to demonstrate the use of $MFSTL$. 

In a Loan Approval System, clients (customers) apply for loans of different sizes and wait for the appropriate response from the server (loan officer). The client with a request  for a particular loan amount can be seen as a client of {\em that} type. Therefore, we can have client types as say, $\Gamma_0=\{h,l,m\}$ and client properties  as $P_c=\{req_h,req_l,ans_h,ans_l,req_m,ans_m\}$. Here $h$ means a loan request of type (size) {\sf high}, $l$ means a loan request of type (size) {\sf low} and $m$ means a loan request of type (size) {\sf medium}. Now, we can write a few simple specifications in $MFSTL$ as follows:
\begin{enumerate}
\item $\psi_0 = \lnot\big((\exists x:h) req_h(x) \lor (\exists x:l)req_l(x)\lor (\exists x:m)req_m(x)\big)$\\
which means {\sf initially there are no pending requests}.
\item $\psi_1= G [(\exists x:l)req_l(x) \supset X (\exists y:l)ans_l(y)]$ \\
which means {\sf whenever there is a request of type {\em low} there is an approval for type {\em low} in the next instant}.
\item $\psi_2= G [(\exists x:h)req_h(x) \supset \lnot(\exists y:l)req_l(y)]$ \\
which means {\sf there is no request of type {\em low} taken up as long as there is a {\em high} request pending}.
\item $\psi_3=  G [(\exists x:l) req_l(x) \lor (\exists y:h)req_h(y)\lor (\exists z:m)req_m(z)]$\\
which means {\sf there is at least one request of each type pending all the time}.
\item $\psi_4= G [(\exists x:h)req_h(x) \supset \lnot[(\exists y:l)req_l(y) \lor (\exists y:l)req_l(y)]]$
which is similar to $\psi_2$, {\sf there are no pending medium or low requests with a high request}.
\end{enumerate}
Note that none of these formulae make use of equality ($=$) predicate. Using $=$, we can make stronger statements as follows:
\begin{enumerate}
\item $\psi_5 =  G [(\exists x:h)req_h(x) \land (\forall y:h) \big(req_h(y) \supset x=y\big)]$\\
which means {\sf at all times there is {\em exactly one} pending request of type {\em high}}.
\item $\psi_6 =  G [\big(\lnot (\exists x:h)req_h(x)\big)\lor \Big((\exists x:h)req_h(x) \land (\forall y:h) \big(req_h(y) \supset x=y\big)\Big)]$\\
which means {\sf at all times there is {\em at most} one pending request of type {\em high}}.
\end{enumerate}
In the same vein, using $=$, we can count the requests of each type and say more interesting things. For example, if $\varphi_h^2=(\exists x:h)(\exists y:h)(\exists z:h)\big(req_h(x)\land req_h(y)\land req_h(z) \supset (x=y \lor y=z)\big)$ asserted at a point means there are at most $2$ requests of type $h$ pending then we can frame the following formula:
\begin{itemize}
\item $\psi_5=  G  (\varphi_h^2 \supset X  (\varphi_h^2 \supset  G  \varphi_h^2))$\\
which means, {\sf if there are at most two pending requests of type {\em high} at successive instants then thereafter the number stabilizes}.
\end{itemize}

Unfortunately, owing to a lack of provision for free variables in the scope of temporal modalities, we can't write specifications which seek to match requests and approvals. Here is a sample.
\[ G  ((\forall x)req_u(x) \supset X   F  ans_u(x))\]
which means, {\sf if there is a request of type $u$ at some point of time then the \textbf{same} is approved some time in future}.

The challenge is to come up with appropriate constraints on specifications which allow us to express interesting properties as well as remain decidable to verify.

\section{Model Checking $MFSTL$ against $SPS$}
For model checking the client-server system is modeled as an SPS, $M$, and 
the specification is given by a formula $\psi_0$ in  $MFSTL$. The  problem is 
to check if the system $M$ satisfies the specification $\psi_0$, denoted by 
$M \models \psi_0$. In order to do this we bound the $SPS$ using $\psi_0$ and 
define an {\sf interpreted} version. 
\subsubsection*{Bounded Interpreted $SPS$}
Let $M=(S,\delta,I)$ be an $SPS$ and $\psi_0$ be a specification in  $MFSTL$. 
From $\psi_0$ we compute $V_i({\psi_0})$,  the variables 
of type $u_i$ occurring in $\psi_0$, for each $u_i \in \Gamma_0$. 
Let $|V_i({\psi_0})|=r_i$. Also, $\psi_0$ can 
have at most $2$ client predicates in it, $req_{u_i}$ and $ans_{u_i}$. 
For each $u_i$, we define the bound $n_i=4\times r_i$. Let 
$\mathbb{N}_i=\{0,1,2,\cdots,n_i\}$ and $P_i=\{pi[j],qi[j]\mid 1\le j \le r_i\}$. 
Further, $\mathbb{P}=\bigcup_{i}2^{P_i}$.

Now, we are in a position to define an interpreted form of bounded SPS.
The interpreted $SPS$ $\mathcal{M}=(\Omega,\Rightarrow,\calI,Val)$ is as follows:
\begin{enumerate}
\item $\Omega = S \times \prod\limits_{i=1}^{k}\mathbb{N}_i\times \mathbb{P}$, where each 
configuration $(s,\sigma,\varrho)\in \Omega$ satisfies the following condition: 
for each $u_i \in \Gamma_0$, for each $1 \le j\le r_i$, $\{pi[j],qi[j]\} \not \subset \varrho$.
\item $\calI= \{(s,<0,0,\cdots,0>,\emptyset)\mid s \in I\}$
\item $Val:\Omega \to 2^{P_s}$
\item $\Rightarrow\subseteq \Omega\times\Gamma\times \Omega$ as follows:
$(s,\sigma,\varrho)\Longstep{r}(s^{\prime},\sigma^{\prime},\varrho^{\prime})$ 
iff $(s,r,s^{\prime})\in \delta$ and the following conditions hold:
\begin{enumerate}
\item when $r=\tau$, $\sigma=\sigma^{\prime}$ and $\varrho=\varrho^{\prime}$.

\item when $r=req_{u_i}$, $\sigma^{\prime}[i]=\sigma[i]+1$ and $\varrho^{\prime}=\varrho\cup\{pi[\sigma^{\prime}[i]]\}$.

\item when $r=ans_{u_i}$, $\sigma^{\prime}[i]=\sigma[i]-1$ and $\varrho^{\prime}=\varrho\setminus\{pi[\sigma[i]]\}\cup\{qi[\sigma[i]]\}$.
\end{enumerate}
\end{enumerate}

In order to use a model checking tool, we need to eliminate quantifiers from $\psi_0$ and
thereby convert it to a standard LTL form. This can be done as follows. Let $\varphi$ be
an $MFO$ sentence of type $u_i$ occurring in $\psi_0$. Convert $\varphi$ into an equivalent
propositional logic formula $\alpha$ using the bound $r_i$ (and model $\{1,2,\cdots,r_i\}$) 
in the standard way \cite{BJ89}. For uniformity, we replace $req_{u_i}(j)$ by $pi[j]$ and 
$ans_{u_i}(j)$ by $qi[j]$, respectively.

Once we have transformed every $MFO$ sentence occurring in $\psi_0$ to equivalent propositional
formula, $\psi_0$ turns out to be an LTL formula $\phi$. Therefore, we can model check
$\mathcal{M}$ against $\phi$ using NuSMV.
Consequently, we can assert the following theorem:
\begin{thm}[\cite{She11}]
$M\models\psi_0$ can be checked in double exponential time.
\end{thm} 
\subsection{Model Checking Using NuSMV}
We check whether a $SPS$-based system $M$ satisfies an $MFSTL$ specification $\psi_0$
as follows. We employ the .smv encoding for describing the $SPS$ and related
$MFSTL$ specification, as in the case of standard NuSMV input, which has CTL/LTL
specifications, instead. The following NuSMV code describes an $SPS$ followed
by an $MFSTL$ formula. Though, the syntax of $SPS$ is standard Kripke, it has
a different multi-counter automata semantics. Therefore, it has to be converted
to standard finite state transition machine along with a translation to LTL
from $MFSTL$ which is accomplished through quantifier elimination of 
MFO-sentences.

For simplicity, we assume that the $SPS$ admits no internal action $\tau$ and
also, only one client type $u_0$. Consequently, the input alphabet $\Gamma$
can be modeled by a single input variable $ip$ in the NuSMV encoding.
Therefore, $ip=TRUE$ encodes $req_0$ and $ip=FALSE$ encodes $ans_0$.
The state set is modeled by the system variable $loc$ which admits
two possible states $s_0$ and $s_1$. Initially the system is in state
$s_0$. When, there is a request in state $s_0$, the system moves to
state $s_1$. When there is a response in state $s_1$, the system
moves to $s_0$. In all other scenarios, the state does not change.
\begin{verbatim}
MODULE main
IVAR ip : boolean;
VAR  loc : {s0,s1};
ASSIGN
 init(loc):=s0;
 next(loc):=case
  loc=s0 & ip=TRUE : s1;
  loc=s1 & ip=FALSE : s0;
  TRUE : loc;
 esac;

MFSTLSPEC
 G((Ex)p(x) -> X((Ex)q(x)))
\end{verbatim}
Along with the NuSMV code there is an $MFSTL$ specification,
$ G((Ex)p(x) \to X((Ex)q(x)))$, where $E$ represents $\exists$ and
$A$ represents $\forall$. Here the predicates $p$ and $q$ encode $req_0$  
and $ans_0$, respectively. The specification means that at all time instances,
if there is a new request then it is followed immediately by a response
in the next instance.

The pseudo-NuSMV input above is converted to standard NuSMV input
as given below. A brief explanation is given alongside the code. There is no
change in the behaviour of $loc$. The modified code has the behaviour of the 
newly added variables, $ctr$, $p[1]\cdots p[4]$ and $q[1]\cdots q[4]$.

We first look at the $MFSTL$ specification and compute the bound from the 
$MFO$ sentences contained therein. Then, the formula is converted to LTL form
by eliminating the quantifiers from the $MFO$ sentences, in the standard way. 
The bound computed above, is used to define limit of the variable $ctr$ in 
the modified SMV code and also the size of the arrays $p$ and $q$. 

In the present case, the bound is $4$ as the number of $MFO$ predicates in 
the original specification is $2$ ($p$ and $q$) and only one variable $x$ is used.
\begin{verbatim}
MODULE main
IVAR  ip : boolean;
VAR   loc : {s0,s1};
VAR   ctr: 0..4;
VAR   p : array 1..4 of boolean;
VAR   q : array 1..4 of boolean;
ASSIGN
 init(loc):=s0;
 init(ctr):=0;
 init(p[1]):=FALSE; 
 init(p[2]):=FALSE;
 init(p[3]):=FALSE; 
 init(p[4]):=FALSE;
 init(q[1]):=FALSE; 
 init(q[2]):=FALSE;
 init(q[3]):=FALSE; 
 init(q[4]):=FALSE;
 next(loc):= case
        loc=s0 & ip=TRUE  : s1;
        loc=s1 & ip=FALSE : s0;
        TRUE		  : loc;
 esac;
--ctr tracks the number of active clients.
--it's value may range from 0 to 4. 
--these values give the client names.
 next(ctr):= case
        ip=TRUE	 & ctr <4 : ctr + 1;
        ip=FALSE & ctr >0 : ctr - 1;
        TRUE		   : ctr;
 esac;
--p[i] is TRUE means request from client i is active
--p[i] is FALSE means request from client i is inactive
 next(p[1]):= case
        p[1]=FALSE & ip=TRUE & ctr=0 : TRUE;
        p[1]=TRUE & ip=FALSE & ctr=1 : FALSE;
        TRUE			     : p[1];
 esac;
 next(p[2]):= case
        p[2]=FALSE & ip=TRUE & ctr=1 : TRUE;
        p[2]=TRUE & ip=FALSE & ctr=2 : FALSE;
        TRUE			     : p[2];
 esac;
 next(p[3]):= case
        p[3]=FALSE & ip=TRUE & ctr=2 : TRUE;
        p[3]=TRUE & ip=FALSE & ctr=3 : FALSE;
        TRUE			     : p[3];
 esac;
 next(p[4]):= case
        p[4]=FALSE & ip=TRUE & ctr=3 : TRUE;
        p[4]=TRUE & ip=FALSE & ctr=4 : FALSE;
        TRUE			     : p[4];
 esac;
--q[i] is TRUE means request from client i is served
--q[i] moves to FALSE in the next time instance.
 next(q[1]):= case
        q[1]=FALSE & ip=FALSE & ctr=1 : TRUE;
        q[1]=TRUE		      : FALSE;	
        TRUE			     : q[1];
 esac;
 next(q[2]):= case
        q[2]=FALSE & ip=FALSE & ctr=2 : TRUE;
        q[2]=TRUE		      : FALSE;
        TRUE			     : q[2];
 esac;
 next(q[3]):= case
        q[3]=FALSE & ip=FALSE & ctr=3 : TRUE;
        q[3]=TRUE		      : FALSE;
        TRUE			     : q[3];
 esac;
 next(q[4]):= case
        q[4]=FALSE & ip=FALSE & ctr=4 : TRUE;
        q[4]=TRUE		      : FALSE;
        TRUE			     : q[4];
 esac;

LTLSPEC        
 G ( (p[1] | p[2] | p[3] | p[4]) -> 
            (q[1] | q[2] | q[3] | q[4]) )
\end{verbatim}
This NuSMV description is sent for model checking using NuSMV, which, accordingly,
decides whether the specification holds true for the $SPS$ system. It is not difficult
to see that this scheme can be easily extended to systems with multiple client types.
The code for this tool is available on bitbucket \url{https://bitbucket.org/ssnmfotl/nusmv-mfotl-code}.

\section{Discussion \& Future Work}
We have presented a simple model for client - server systems where the number
of clients is known only at run-time and hence unbounded. The model is 
equivalent to multi-counter automata and so we can decide reachability
properties. We also proposed a simple temporal logic over sentences of
a monadic first-order logic. We proposed a practical scheme to implement
model checking such client - server systems against the given logic
specifications.

The immediate challenge is to extend the theory to include active clients, where
there is non-trivial interaction between the client and server between the send-request
and the request-response. We call these client - server systems with active clientele.
However, for such systems we need specifications that admit temporal modalities in the 
scope of quantifiers, and controlling the expressiveness while retaining decidability 
is challenging.

\bibliographystyle{unsrt}

\end{document}